\documentclass[iop]{emulateapj}

\newcommand{\src}{PSR~J1622$-$4950}
\newcommand{\chandra}{\textit{Chandra}}
\newcommand{\xmm}{\textit{XMM-Newton}}
\newcommand{\swift}{\textit{Swift}}
\newcommand{\tempotwo}{{\tt{TEMPO2}}}
\newcommand{\firstradio}{XTE~J1810$-$197}
\newcommand{\fift}{1E~1547.0$-$5408}
\newcommand{\tenfour}{1E~1048.1$-$5937}
\newcommand{\sgrsgr}{SGR~J1745$-$2900}

\usepackage{amsmath}
\usepackage[colorlinks,allcolors=blue]{hyperref}

\begin{document}

\title{Spin-down evolution and radio disappearance of the magnetar \src}

\author{P.~Scholz\altaffilmark{1,2},
F.~Camilo\altaffilmark{3},
J.~Sarkissian\altaffilmark{4},
J.~E.~Reynolds\altaffilmark{5},
L.~Levin\altaffilmark{6},
M.~Bailes\altaffilmark{7,8},
M.~Burgay\altaffilmark{9}
S.~Johnston\altaffilmark{5},
M.~Kramer\altaffilmark{10,6},
and
A.~Possenti\altaffilmark{9}}
\altaffiltext{1}{National Research Council of Canada, 
Herzberg Astronomy and Astrophysics, Dominion Radio Astrophysical Observatory,
P.O. Box 248, Penticton, BC V2A 6J9, Canada; 
\email{paul.scholz@nrc-cnrc.gc.ca}}
\altaffiltext{2}{Dept. of Physics and McGill Space Institute, Rutherford Physics Building,
McGill University, 3600 University Street, Montreal, Quebec, H3A
2T8, Canada}
\altaffiltext{3}{SKA South Africa, Pinelands, 7405, South Africa} 
\altaffiltext{4}{CSIRO Parkes Observatory, Parkes, NSW 2870, Australia}
\altaffiltext{5}{CSIRO Astronomy and Space Science, Australia
Telescope National Facility, Epping, NSW 1710, Australia}
\altaffiltext{6}{Jodrell Bank Centre for Astrophysics, School of
Physics and Astronomy, The University of Manchester, Manchester M13
9PL, UK}
\altaffiltext{7}{Centre for Astrophysics and Supercomputing, 
Swinburne University of Technology, Mail H30, PO Box 218, Hawthorn, VIC 3122, Australia} 
\altaffiltext{8}{ARC Centre of Excellence for All-Sky Astronomy (CAASTRO)} 
\altaffiltext{9}{INAF -- Osservatorio Astronomico di Cagliari, 
Via della Scienza 5, I-09047 Selargius (CA), Italy}
\altaffiltext{10}{Max-Planck-Institut f\"ur Radioastronomie, Auf dem H\"ugel 
69, D-53121 Bonn, Germany}

\begin{abstract}
We report on 2.4\,yr of radio timing measurements of the magnetar
\src\ using the Parkes telescope, between 2011 November and 2014
March.  During this period the torque on the neutron star (inferred
from the rotational frequency derivative) varied greatly, though
much less erratically than in the 2\,yr following its discovery in
2009.  During the last year of our measurements the frequency
derivative decreased in magnitude monotonically by 20\%, to a value
of $-1.3\times10^{-13}$\,s$^{-2}$, a factor of 8 smaller than when
discovered. The flux density continued to vary greatly during our
monitoring through 2014 March, reaching a relatively steady low
level after late 2012.  The pulse profile varied secularly on a
similar timescale as the flux density and torque. A relatively rapid
transition in all three properties is evident in early 2013. After
\src\ was detected in all of our 87 observations up to 2014 March,
we did not detect the magnetar in our resumed monitoring starting
in 2015 January and have not detected it in any of the 30 observations
done through 2016 September.
\end{abstract}


\section{Introduction}

Magnetars are a class of neutron stars with extremely high magnetic
fields ($B\sim10^{13-15}$\,G) and long spin periods (2--12\,s).
Their high-energy emission is powered via decay of their magnetic
fields, rather than through rotation.  This is revealed through
large outbursts and X-ray luminosities that exceed the available
rotational spin-down luminosity \citep[for reviews see][]{wt06,mer08}.
During outburst, magnetars can increase their X-ray fluxes by orders
of magnitude, and then fade on a timescale of months to years.

Most magnetars have been discovered and monitored in X-rays. The
best characterized are those that have been monitored for over
15 years with the {\em Rossi X-ray Timing Explorer} \citep{dk14},
now continued by \swift\ \citep[e.g.,][]{akn+12,akn+15}.  This may
be a biased sample, as only five of the 23 known magnetars
\citep{ok14}\footnote{see the online Magnetar Catalog at \url{http://www.physics.mcgill.ca/~pulsar/magnetar/main.html}}
are persistently bright enough to be monitored in this way \citep{dk14}.  In order
to expand our understanding of magnetars it is desirable to perform
detailed, long-term monitoring of the rotational and radiative
behavior of more objects.

Radio emission has been detected from only four of the known
magnetars, but it is often quite bright \citep[e.g.,][]{crhr07,sj13}.
We can therefore expand the sample of well characterized magnetars
by performing long-term monitoring using radio telescopes. The study
of radio emission from magnetars also provides a new electromagnetic
window into the behavior of these most magnetic objects known.

\firstradio\ was the first magnetar to be detected in radio
\citep{crh+06}, followed shortly by \fift\ \citep{crhr07}. They
were found to have highly variable radio flux densities and pulse
profiles, unlike ordinary pulsars. Very unusually, their radio
spectra are generally flat \citep[e.g.,][]{crp+07}. 
Both are transient radio 
sources: radio emission from \firstradio\ turned off in 2008 \citep{crh+16}
and \fift\ was detected intermittently following its 2009 outburst 
\citep{chr09,bip+09}.
The third radio
magnetar, \src, is the subject of this work. A fourth was more recently
discovered 2\arcsec\ away from the Galactic center and is the closest
known pulsar to Sgr~A* \citep{efk+13,sj13}.

\src\ was discovered with the CSIRO Parkes telescope as a radio
pulsar with period $P=4.3$\,s and dispersion measure $\mbox{DM} =
820$\,pc\,cm$^{-3}$ \citep{lbb+10}.  To date it remains the only
magnetar to have been detected in the radio without prior knowledge
of a corresponding X-ray source.  Like other radio magnetars it has
a flat spectrum, nearly 100\% linear polarization, and highly variable 
flux density and pulse profiles.
Its rotational behavior following discovery was characterized by
Parkes observations between 2009 April and 2011 February \citep{lbb+12}.
Long-term phase-connected timing solutions were not possible due
to the rapidly evolving spin-frequency derivative, $\dot \nu$, and
insufficient observing cadence.  From short-term timing solutions,
$|\dot \nu|$ was found to have decreased by a factor of
2 in the 2\,yr following discovery.

\src\ was identified as an X-ray source using archival and
dedicated \chandra\ and \xmm\ observations.  Its X-ray flux decreased
by a factor of $\sim$50 between 2007 June and 2011 February,
presumably following a pre-discovery outburst.  X-ray pulsations
have not been detected, implying a 70\% limit on the pulsed fraction
\citep{ags+12}.

Here we present the analysis and results of an additional 2.4\,yr
of Parkes observations of \src. We describe our dataset in
Section~\ref{sec:obs}.  In Sections~\ref{sec:profile} and \ref{sec:flux}
we show the pulse profile and flux density evolution of the source.
In Section~\ref{sec:timing} we present a timing analysis and the
resulting phase-connected timing solutions.  We discuss our results
in Section~\ref{sec:disc}, and conclude in Section~\ref{sec:conc}.

\begin{figure*}
\plotone{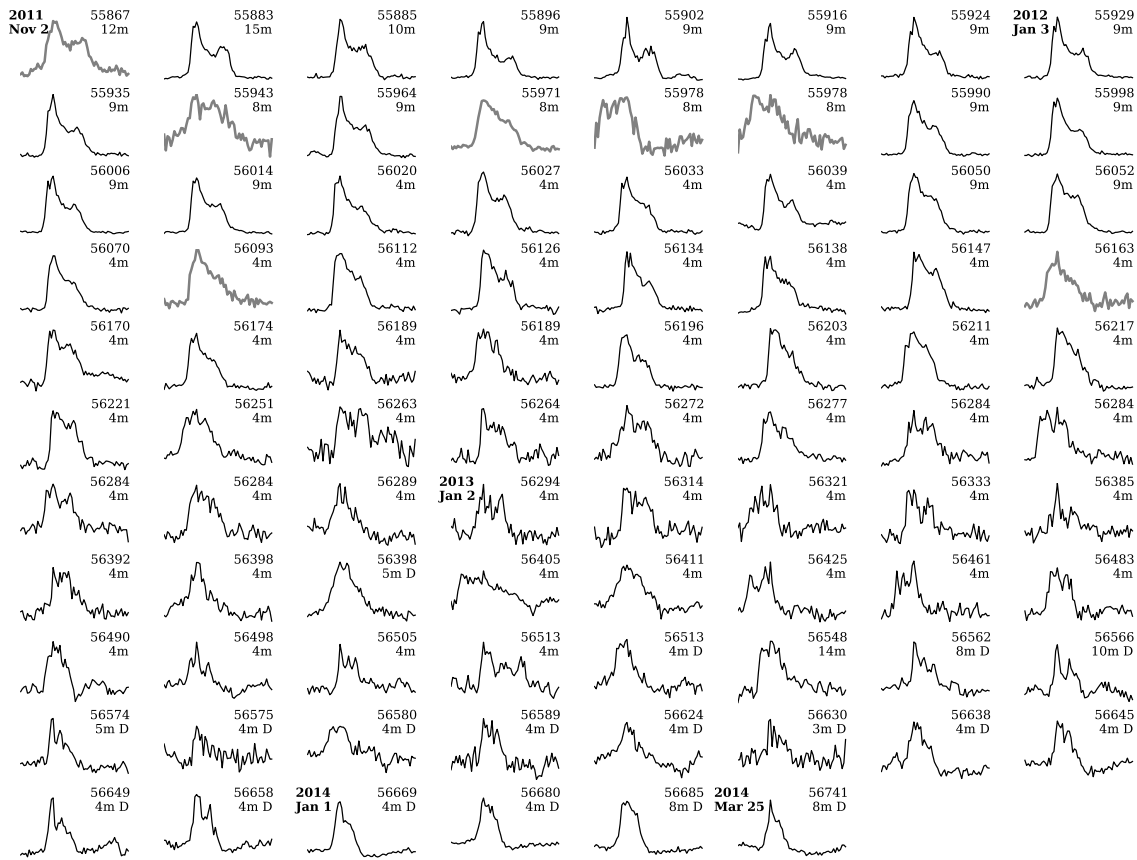}
\figcaption{
Radio pulse profiles of \src.  Black (gray) profiles correspond
to 3\,GHz (1.4\,GHz) observations. The full pulse period of 4.3\,s
is displayed, with 64 phase bins, and profiles are arbitrarily
aligned.  We list the MJD and integration time (in minutes) of each
observation, along with calendar dates in select instances.  Those
observations that used a digital filterbank are denoted by a ``D''.
All other profiles, obtained with an analog filterbank, have been
corrected to account for instrumental artifacts (see
Section~\ref{sec:profile}).  Some profiles remain somewhat contaminated
by RFI.
\label{fig:allprofs}
}
\end{figure*}

\section{Observations}
\label{sec:obs}

We observed \src\ at Parkes on a regular basis between 2011 November
and 2014 March. These observations were typically done on the same
days when we monitored the magnetar 1E~1547.0$-$5408, which was
largely observed at frequencies near 3\,GHz because severe scattering
renders its pulse hard to detect at 1.4\,GHz \citep{crhr07}.  We
did a total of 87 observations on 81 days, 90\% of them at 3\,GHz
using the 10-50\,cm receiver, and the remainder at 1.4\,GHz using
the center beam of the 20\,cm multibeam receiver \citep{swb+96},
once every 10 days on average.  Integration times were typically 5
or 10 minutes per observation.

A total of 69 observations through 2013 September were done with
the analog filterbank system \citep[AFB; see, e.g.,][]{mlc+01},
used to sample a bandwidth of 864\,MHz centered on 3078\,MHz, or a
288\,MHz band centered on 1374\,MHz. In each case the individual
channel widths were 3\,MHz.  In 2013 April we began using PDFB3, a
digital filterbank (DFB), centered on 3100\,MHz to sample 512
2\,MHz-wide channels. In all cases we recorded total-intensity
(polarization-summed) search-mode data using 1\,ms samples.

Each of the datasets was subsequently dedispersed and folded using
a known ephemeris \citep{lbb+10}.  Each folded observation was
inspected for frequency channels and sub-integrations that were
highly contaminated by radio frequency interference (RFI). The
contaminated channels and sub-integrations were then masked in all
subsequent analysis.

\src\ was detected in every observation we did during 2011--2014.
We resumed observations on 2015 January 11, but have not detected
the pulsar in any of 30 epochs through 2016 September 16. These observations,
largely at 3\,GHz using the PDFB4 digital filterbank, lasted for
15 minutes on average.

In addition to the new observations done between 2011 and 2014, we
use the flux densities and pulse times-of-arrival (TOAs) reported
in \citet{lbb+12} from Parkes observations between 2009 and 2011.
We also utilize 26 archival observations done between the dataset
presented in \citet{lbb+12} and the beginning of our campaign. These
observations included 15 observations at 1.4\,GHz and three at
3.1\,GHz using PDFB3/4, and eight 1.4\,GHz observations with the
CASPER-Parkes-Swinburne Recorder (CASPSR).

\section{Analysis and Results}

\begin{figure}
\plotone{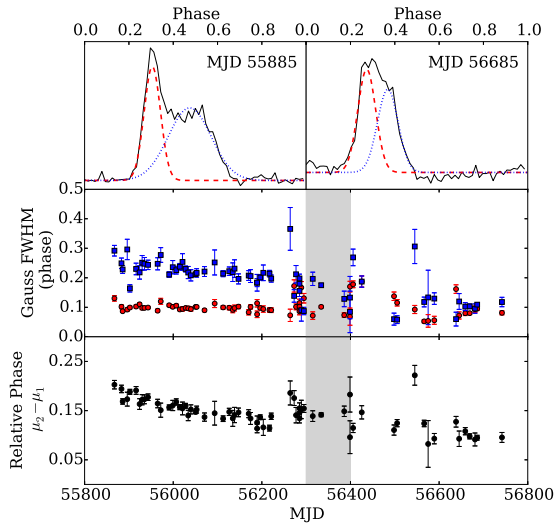}
\figcaption{
Gaussian fits to pulse profiles of \src.  Top: Two example profiles
are shown with their two-component Gaussian fits shown.  Middle:
The FWHM of the two Gaussian components for each profile.  Bottom:
The separation between the peak phases of the two Gaussian components
for each profile.  The grey bar marks the period between MJD 56300
and 56400 in order to illustrate the correlated behavior between
the spin-down (Figure~\ref{fig:resids}), flux density
(Figure~\ref{fig:fluxes}), and pulse profile evolution at that time.
\label{fig:gauss_fits}
}
\end{figure}

\subsection{Pulse profile variations}
\label{sec:profile}

Similar to what was observed between 2009 and 2011 by \citet{lbb+12},
the pulse profile of \src\ in our observations is made up of multiple
components that vary in relative amplitude and separation over time.
Figure~\ref{fig:allprofs} shows the profiles for all of our 2011--2014
observations.  Long-period pulsars observed with the AFB system
display artifacts caused by a high-pass filter with $\approx 0.9$\,s
time constant.  We used the prescription given by \citet{mlc+01}
to correct for this effect in the profiles presented in
Figure~\ref{fig:allprofs}.

In late 2011 the profiles were clearly composed of two peaks, with
the second fainter than the first. In late 2012 the pulsar became
significantly fainter (see Section~\ref{sec:flux}) and more affected
by RFI.  Often the profile could only be resolved as a broad single
peak. This persisted until late 2013 when the flux density increased
slightly and the pulse profile narrowed
(Figure~\ref{fig:allprofs}).

To quantify the narrowing of the pulse profile, we fit a two-Gaussian model
to the profiles. The model fit to the profiles is
\begin{equation}
P(\phi,A_i,\mu_i,\sigma_i) =
A_1 \exp{\frac{-(\phi-\mu_1)^2}{2\sigma_1^2}} +
A_2 \exp{\frac{-(\phi-\mu_2)^2}{2\sigma_2^2}},
\end{equation}
where $A_i$ are the amplitudes, $\mu_i$ are the peak phases, and
$\sigma_i$ are the widths of the Gaussian components. The full-width
at half-maximum (FWHM) of the components are $2\sqrt{2\ln2} \sigma_i$.
The results of these fits are shown
in Figure \ref{fig:gauss_fits} and clear evolution is evident.  The
leading component remained relatively constant in width and the
trailing Gaussian component became narrower and closer in phase
to the first component as a function of time.  This change was
occurring on a similar timescale to the radio flux density decrease
(see Section \ref{sec:flux}).

\begin{figure*}
\plotone{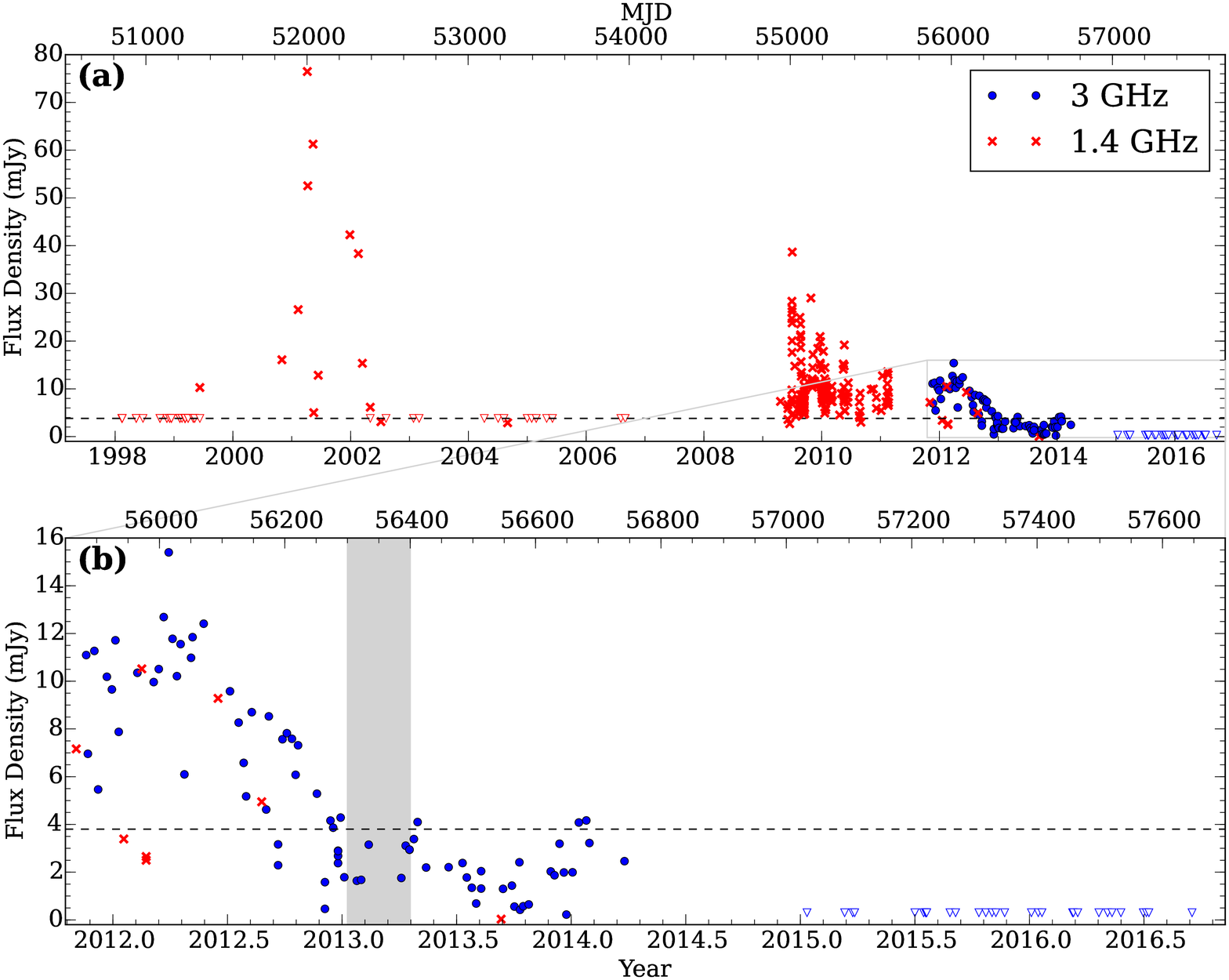}
\figcaption{Top: Flux density for each detection of \src.  Bottom:
Zoom-in on the flux density measurements from the new data presented
in this work.  Red crosses represent 1.4\,GHz observations and blue
circles represent 3\,GHz observations. Open red triangles represent
non-detections from the off-axis 1.4\,GHz pre-discovery observations 
\citep{lbb+10}. They have a limiting flux density of 3.8\,mJy (see text),
indicated by the dashed line.  Non-detections starting in early
2015, largely at 3\,GHz with a limiting flux density of 0.3\,mJy,
are shown as downward-pointing blue triangles.  Measurements between 2009 and
mid 2011 are those from \citet{lbb+12} multiplied by a scaling factor
of 2.3 (see text).  The grey bar marks the
period between MJD 56300 and 56400 in order to illustrate the
correlated behavior between the spin-down (Figure~\ref{fig:resids}),
flux density, and pulse profile evolution (Figure~\ref{fig:gauss_fits})
at that time. \label{fig:fluxes}}
\end{figure*}

\subsection{Flux density evolution}
\label{sec:flux}

Our filterbank data were not flux calibrated. Nevertheless, we can
extract useful flux density measurements by computing the area under
each profile and scaling it to a Jansky scale using the system
equivalent flux density (SEFD) at the location of the pulsar. 
First we
set the off-pulse level to zero and normalized the summed pulse
profile counts by the off-pulse rms.  We then scaled the profile
into units of flux density using the off-pulse rms from the radiometer
equation \citep{dtws85}:
\begin{equation} 
\frac{ \beta\, \mathrm{SEFD} }{\sqrt{n_p t_{\mathrm{int}} \Delta f }},
\end{equation} 
where $\beta$ is a loss factor
due to digitization of the signal (1.5 for the AFB, 1.1 for the
DFB), $n_p=2$ is the number of polarizations summed, $t_{\mathrm{int}}$
is the integration time per phase bin, and $\Delta f$ is the
bandwidth.  
We determined the SEFD by analyzing with PSRCHIVE \citep{hvm04}
full-Stokes calibrated observations done with PDFB3. The SEFD at
3.1\,GHz is 61\,Jy (based on an observation done on 2011 December
12), while the SEFD at 1.4\,GHz is 69\,Jy (based on an observation
done on 2010 November 3), both measured with about 5\% precision.
Flux densities measured in this way are shown in
Figure~\ref{fig:fluxes}. In the absence of residual RFI and other
profile artifacts, we estimate that the absolute fractional uncertainty
for each measurement is $\approx 25$\%. However, some profiles were
contaminated by residual RFI (see Figure~\ref{fig:allprofs}). To
address this, we made the measurements using two independent tools:
one in which the off-pulse regions are chosen arbitrarily by a user
and one in which the off-pulse regions are determined automatically
by growing the off-pulse region until the variance of the off-pulse
data changes by more than 10\%.
These two tools yield different off-pulse baseline estimations, 
normalizations, and flux density values. While on occasion the two
measures differed by up to $\approx 50$\%, in most cases they agreed
more closely and these discrepancies do not affect the trends visible
in Figure~\ref{fig:fluxes}.

\begin{deluxetable*}{lcc}
\tablecaption{Two timing solutions for \src\
\label{ta:timing}}
\tablewidth{0.6\linewidth}
\tablehead{
\colhead{Parameter}&\colhead{Solution 1}&\colhead{Solution 2}
}
\startdata
\multicolumn{3}{c}{Timing Parameters} \\                                                                       
[1pt]
\hline \\
[-6pt]
Right ascension (J2000)\tablenotemark{a}                & 16:22:44.8           & 16:22:44.8         \\ 
Declination (J2000)\tablenotemark{a}                    & $-$49:50:54.4        & $-$49:50:54.4        \\ 
Dispersion measure, DM (pc\,cm$^{-3}$)\tablenotemark{a} & 820                     & 820                     \\ 
Spin frequency, $\nu$ (s$^{-1}$)                        & $0.231115433(4)$           & $0.2311059204(8)$          \\
Frequency derivative, $\dot{\nu}$ (s$^{-2}$)            & $-3.566(3)\times10^{-13}$  & $-1.4827(7)\times10^{-13}$  \\ 
Frequency second derivative, $\ddot{\nu}$ (s$^{-3}$)    & $3.22(9)\times10^{-21}$    & $1.09(3)\times10^{-21}$     \\ 
Epoch of frequency (MJD)                                      & 56100.0                   & 56563.0                   \\ 
Data span (MJD)                                          & 55867--56334               & 56385--56742 \\
Number of TOAs                                          & 55                         & 31                       \\ 
rms residual (phase)                          & 0.12                       & 0.014                \\
\hline             \\
[-6pt]                                                                                             
\multicolumn{3}{c}{Derived Parameters} \\                                                                       
[1pt] 
\hline \\
[-6pt]
Surface dipolar magnetic field, $B$ (G)                 & $1.7\times10^{14}$   & $1.1\times10^{14}$ \\ 
Spin-down luminosity, $\dot{E}$ (erg\,s$^{-1}$)         & $3.2\times10^{33}$    & $1.4\times10^{33}$ \\ 
Characteristic age, $\tau_c$ (kyr)                      & $10$                & $25$  \\
[-5pt] 
\enddata
\tablecomments{Numbers in parentheses are \tempotwo\ $1\,\sigma$
uncertainties.} 
\tablenotetext{a}{Values fixed to those from \citet{lbb+10}.}
\end{deluxetable*}

Following the \src\ discovery, \citet{lbb+10} realized that
pre-discovery observations existed for the years 1998--2006 in the
form of archival search-mode data for two nearby pulsars:
PSR~J1623$-$4949 (11\arcmin\ away) and PSR~J1622$-$4944 (7\arcmin\
away). When pointing at the latter, \src\ was near the half-power
point of the Parkes 1.4\,GHz primary beam ($\mbox{FWHM} = 14.4\arcmin$),
with a reduction in sensitivity by a factor of 1.8. When pointing
at the former, assuming a Gaussian beam (which may not be appropriate
so far off boresight) results in sensitivity reduced by a factor
of 6.3. Astonishingly, \citet{lbb+10} recovered many bright detections
of \src\ even that far off axis, and estimated flux densities. We
have reanalyzed those data (52 individual AFB observations) as
described above for our new dataset in order to place both sets of
detections on the same flux density scale. We made 14 detections
in pre-discovery data, the same as \citet{lbb+10}. 

In Figure~\ref{fig:fluxes} we also include the flux densities
corresponding to the data presented in \citet{lbb+10}. However,
\citet{lbb+10} used $\mbox{SEFD} = T_{\rm sys}/\mbox{Gain} =
24\,\mbox{K}/0.735\,\mbox{K}\,\mbox{Jy}^{-1} = 33$\,Jy, which is a
factor of 2.1 less than our measured value of 69\,Jy at 1.4\,GHz.
Also, they used a loss factor $\beta = 1.0$, while $\beta = 1.1$
for DFB data and 1.5 for AFB data. Thus, we multiply the flux density
values presented in \citet{lbb+10} by 2.3 for DFB data (obtained
during 2009--2011) and 3.1 for AFB data (prior to 2007). Our SEFD
was measured at the position of \src, and the correction factor for
the pre-2007 data (during which the telescope was pointed several
arcminutes away from \src) is therefore uncertain.  Nevertheless,
we judge that our SEFD is a closer approximation to the true value
than the cold-sky value assumed in \citet{lbb+10}.

All flux density measurements (corrected where necessary as described above) 
are summarized in Figure~\ref{fig:fluxes}.
Many of the pre-discovery values are much larger, as well as more
variable, than the more recent ones.  Another notable point is that
we have numerous detections in 2012--2014 with flux densities below the
(corrected) 3.8\,mJy detection limit of the off-axis observations (\citealt{lbb+10}
used a limit of 1.2\,mJy).
Therefore, it is quite possible that some non-detections
for the 1998--2006 period \citep{lbb+10} simply reflect a lack of
sensitivity, and that the pulsar would have been detected had it
been observed on-axis. Those non-detections hence do not necessarily
imply a turnoff in radio emission. By contrast, our consistent
non-detections starting in 2015 (Section~\ref{sec:obs}) reflect a
different state compared to 2009--2014. For the first time in the
study of \src, we can entertain the possibility that the radio
emission effectively turned off or at least transitioned to a significantly
fainter state.

\subsection{Phase-coherent timing}
\label{sec:timing}

In principle, timing of radio magnetars presents particular challenges
owing to the varying pulse profiles. In practice, for \src\ this
did not present substantial difficulties for the post-2011 data
used in this paper.

To account for coarse changes in pulse profiles, we used three
separate templates for TOA extraction.  For
observations prior to MJD~56250, we used the profile observed on
MJD~55924 as the template. For AFB observations from MJD~56250
onwards, the MJD~56284 profile was used.  Finally, the MJD~56685
profile was used to extract TOAs from all DFB observations (see
Figure~\ref{fig:allprofs}).  All TOAs were obtained with the {\tt
PRESTO} \citep{rem02} tool {\tt get\_TOAs.py}.

In order to quantify the effect of the evolving profiles on the
accuracy of the TOAs, we also extracted TOAs using templates built
from multi-Gaussian fits to the profiles observed on MJD~55924 (for
AFB data) and MJD~56669 (for DFB data).  We then measured the
difference between corresponding original TOAs and Gaussian-template
TOAs. The standard deviation of these differences ranged over
20--40\,ms for AFB TOAs and was 30\,ms for the DFB TOAs, i.e., about
1\% of the pulse period.  We added these standard deviations in
quadrature to our nominal TOA uncertainties, to account for the
error introduced in timing the pulsar with a restricted set of
templates in the face of varying pulse profiles.
One observation, on MJD 56545, was too faint to provide
a reliable TOA.

The TOAs were fit to a timing model describing the pulsar rotation
where the pulse phase as a function of time is described by a Taylor
series expansion.  Initially, only the spin frequency $\nu = 1/P$
was fit for, to a set of four TOAs extracted from each observation.
A frequency derivative $\dot\nu$ was estimated from those measurements,
and was used as a starting point for the iterative process of
long-term phase-connection using \tempotwo\ \citep{hem06}.  For the
final fits we extracted one TOA per observation, in order to improve
parameter precision.

\begin{figure*}
\plotone{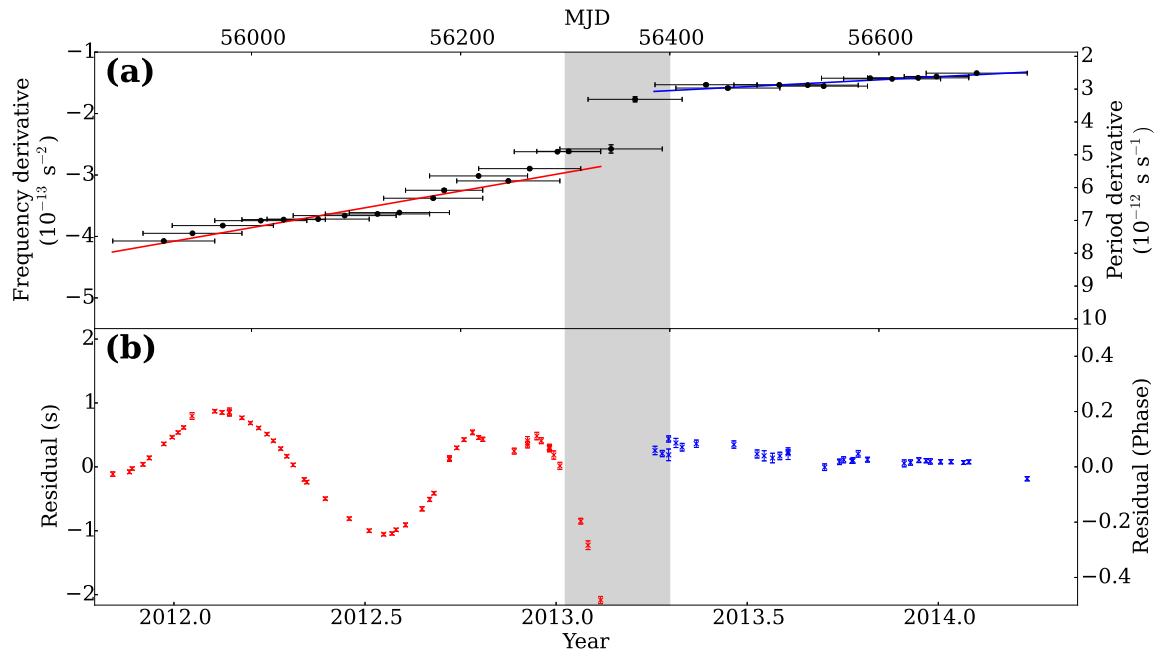}
\figcaption{
The rotation of \src.  Red and blue lines and points correspond to
two simple ($\nu$, $\dot\nu$, $\ddot\nu$) timing solutions.  {\bf
(a)} Frequency derivative.  Black points represent short-term
measurements of the frequency derivative (see Section~\ref{sec:timing}).
{\bf (b)} Timing residuals for the two
timing solutions (Table~\ref{ta:timing}).
The grey bar marks the period between MJD 56300 and 56400 in order
to illustrate the correlated behavior between the spin-down, flux
density (Figure~\ref{fig:fluxes}), and pulse profile evolution
(Figure~\ref{fig:gauss_fits}) at that time.  \label{fig:resids}
}
\end{figure*}

Using simple timing models with only $\nu$, $\dot\nu$,
and $\ddot\nu$, it is possible to phase-connect the dataset
in two separate date ranges.  These solutions are shown in
Table~\ref{ta:timing}.  In Figure~\ref{fig:resids}(a) and (b), the
$\dot\nu$ evolution and phase residuals of these two solutions are
shown in red and blue.  In order to probe the evolution of $\dot
\nu$ in more detail, we also fit short-term overlapping timing
models using only $\nu$ and $\dot\nu$. Each short-term model was
fit over a minimum of five observations spanning a minimum of 61
days and a maximum of 100 days. The resulting values of $\dot \nu$
are shown in Figure~\ref{fig:resids}(a), where the horizontal bars
represent the time span of the fits.

Formal pulsar TOA uncertainties obtained from cross-correlating
observed profiles with templates are often somewhat underestimated.
It is therefore standard practice to increase the TOA errors by a
scaling error factor (EFAC) that yields a reduced $\chi^2 \equiv
1$, ensuring more realistic parameter uncertainties.  We determined
that for our dataset $\mbox{EFAC} = 1.3$, by considering short-term
timing solutions where the effects of ``timing noise'' are negligible.

To probe the timing evolution between the end of the dataset from 
\citet{lbb+12} and the beginning of our 2011--2014 campaign, we also 
extracted TOAs from 26 archival observations (Section \ref{sec:obs}) 
using PSRCHIVE's {\tt pat} utility. We fit timing solutions with 
$\nu$ and $\dot\nu$ to these data in two time spans where phase 
connection was possible. These frequency-derivative measurements are 
shown as red crosses in Figure \ref{fig:fdot_disc}.

\section{Discussion}
\label{sec:disc}

\src\ is the only magnetar whose rotation has been studied exclusively
at radio wavelengths. Much of what we know about its radiative
behavior also relies on radio observations, but importantly we know
that its X-ray flux decreased by a factor of 50 between mid-2007
and early 2011, with an exponential timescale of 1\,yr, following
a presumed earlier outburst \citep{ags+12}. The high-cadence Parkes
monitoring observations that we have presented here, along with
previously published radio results \citep{lbb+10,lbb+12}, allow us
to consider the evolution of \src\ over many years and to place it
in the context of other magnetars.

\subsection{Pulse profile variations}
\label{sec:prof_disc}

The variability of the radio pulse profiles that we observed for
\src\ between late 2011 and early 2014 (Section~\ref{sec:profile}
and Figure~\ref{fig:allprofs}) seems broadly comparable to that
previously reported. However, \citet{lbb+10} present rapidly changing
pulse profiles from 2009 for which we have no counterparts (see
their Figure~1) whereas the pulse profile variation that we observe 
is much smoother (Figures \ref{fig:allprofs} and \ref{fig:gauss_fits}).
It is possible that in this regard the
magnetosphere of \src\ was more unsettled around the earlier time, which
encompassed an epoch of rapidly decreasing X-ray flux \citep{ags+12}.

Correlated behavior between profile variations and spin-down state
is observed in several young pulsars \citep[e.g.][]{lhk+10,ksj13}.
For these pulsars the spin-down torque is correlated with changes
in profile shape quantified by profile width \citep{lhk+10} or
relative component heights \citep{ksj13}.  For \src\ we see a
correlation that may be broadly similar: as the spin-down torque
decreases between late 2011 and early 2014, the width of the profile
decreases as the second Gaussian component narrows and approaches
the first component. We note however that this appears to be a
continuous evolution, unlike correlated behavior between discrete
spin-down and profile states thus far discerned in some ordinary
radio pulsars.

A similar secular decrease in the width of the pulse profile is not 
evident in other radio magnetars. \citet{crh+16} find that for 
\firstradio\ the profiles varied greatly right up to the disappearance of 
radio emission in late 2008 with no obvious secular evolution. 
Following its 2013 outburst, the magnetar \sgrsgr\ showed a widening of
its profile that stabilized after $\sim100$\,days \citep{laks15}.

\subsection{Flux density evolution}
\label{sec:flux_disc}

The flux density variability of \src\ is interesting, especially
when compared to that of two other radio magnetars.  Following its
discovery in 2009, \src\ had a highly variable radio flux density
(ranging over $\sim3-40$\,mJy at 1.4\,GHz) that, on average, appeared
to be somewhat on a downward trend through early 2011 
\citep[Figure \ref{fig:fluxes} and][]{lbb+12}.
Our measurements until late 2012 (largely at 3\,GHz) appear to be
consistent with this description (Figure~\ref{fig:fluxes}).
Thereafter, however, a new regime took hold. Throughout 2013 and
into early 2014, the measured flux density never exceeded 5\,mJy,
and in the latter half of 2013 was mostly below 2\,mJy. Then,
sometime between 2014 March and 2015 January, radio emission from
\src\ ceased (or at least never rose above a flux density of $\approx
0.3$\,mJy). It remained in this state as of 2016 September.

This behavior is reminiscent of that for the first radio magnetar,
\firstradio. Three years after its X-ray outburst and discovery
\citep{ims+04}, the radio flux density was large, fluctuating greatly
on a daily timescale, and generally on a downward trend
\citep[the radio light curve for the three years following the X-ray
outburst is essentially unknown;][]{ccr+07}. After one year at a low average
flux density, but still fluctuating greatly on daily timescales,
the radio emission from \firstradio\ turned off in late 2008 and
has not recurred \citep{crh+16}.

The flux density behavior of \src\ prior to its discovery in 2009,
during the years 1998--2006, may have had a different character.
It appears that the fluctuations might have been larger then, and
more frequent, with no clearly discernible trends \citep[see
Figure~\ref{fig:fluxes}, Section~\ref{sec:flux}, and Figure~1
of][]{lbb+10}. Most of those early detections are from 2000--2002,
with many non-detections in 1998--2000 and 2002--2006. However, the
flux density limits in those years are above the flux densities for
most of our detections in 2013 (Figure~\ref{fig:fluxes}). We therefore
have a range of possible interpretations spanning two extrema: (i)
the radio pulsar did indeed turn on and off multiple times during
1998--2006, sometimes possibly on rapid timescales; (ii) the radio
pulsar was always on since at least 1998, although often in a faint
state, below 3.8\,mJy (as during 2013), until it finally turned off
in 2014.

In this regard it is instructive to consider the behavior of the
second radio magnetar, \fift. When discovered in 2007, it was a
bright and fluctuating radio source, consistently detectable, with
X-ray flux decaying from a presumed prior outburst \citep{crj+08}.
However, following two large X-ray outbursts in 2008 and 2009, radio
pulses became detectable only sporadically \citep{chr09,bip+09}, with
months-long periods of no detection interspersed with sometimes
hugely bright emission --- all the while the X-ray flux decaying
very slowly (F.\ Camilo et al., in preparation).

The current radio state of \src, and its reduced flux density for
one year prior to turning off (Figure~\ref{fig:fluxes}), more closely
resemble the behavior of \firstradio\ than that so far displayed
by \fift. This is particularly the case when considering the parallel
behavior of the spin-down torque.

\subsection{Spin-down behavior}

\begin{figure}
\includegraphics[width=\columnwidth]{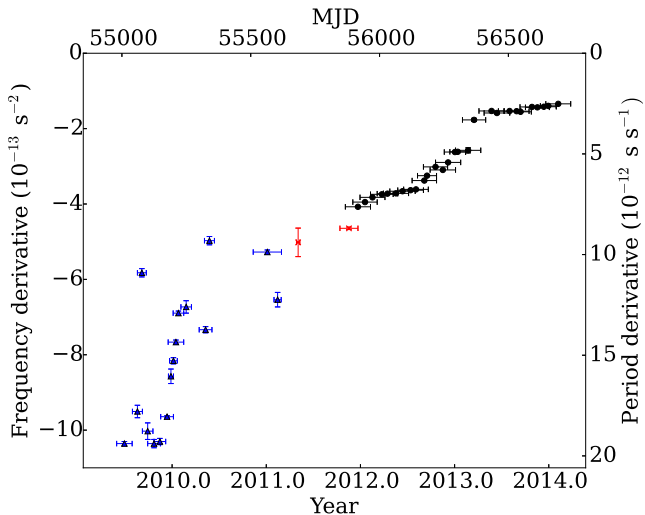}
\figcaption{
The frequency derivative of \src\ from discovery until radio
disappearance. Measurements through early 2011, shown as blue triangles,
 are reproduced from \citet{lbb+12}. Red crosses show frequency-derivative
measurements derived from additional archival observations (Section \ref{sec:timing});
the second measurement was obtained using a combination of archival
data and our initial 2011 observations. Frequency derivative measurements
 from our 2011--2014 campaign (Figure~\ref{fig:resids}a) are 
represented by black circles. 
\label{fig:fdot_disc}
}
\end{figure}

The known spin-down history of \src\ is summarized in
Figure~\ref{fig:fdot_disc}, where we reproduce the measurements of
\citet{lbb+12} for 2009--2011 and add our own for 2011--2014 (see
Figure~\ref{fig:resids}a and Section~\ref{sec:timing}). The detections
in 1999--2003 (Figure~\ref{fig:fluxes}a) are too sparse to yield
$\dot \nu$ measurements.

We identify two regimes: for 2\,yr until mid-2011, the torque on
the neutron star (proportional to $|\dot \nu|$) varied erratically,
both increasing and decreasing, and at very large rates; for the
2.4\,yr until early 2014, the torque decreased monotonically, and
generally at a lower rate than before. Overall, from 2009 until
2014, the torque decreased by a factor of 8.

While the torque that we measured during 2011--2014 decreased
monotonically, it did not do so at a constant rate
(Figure~\ref{fig:resids}a). Until 2013 March, $|\dot \nu|$ decreased
at a rate $\ddot\nu=3\times10^{-21}$\,s$^{-3}$. Then, within
approximately one month, the torque decreased by almost a factor
of 2.  Following this, its rate of change dropped markedly to
$\ddot\nu=1\times10^{-21}$\,s$^{-3}$ (Table~\ref{ta:timing}).
Interestingly, early 2013 is also when the flux density plateaued
at a low level (Figure~\ref{fig:fluxes}b and
Section~\ref{sec:flux_disc}).  In the last year of radio emission,
$|\dot \nu|$ decreased by 20\%, at a rate nearly one order of
magnitude below the average for 2009--2014.

Such torque behavior is not unprecedented in magnetars. Following
X-ray outbursts in 2002, 2007, and 2012, \tenfour\ displayed episodes
where its $\dot\nu$ both increased and decreased repeatedly by up
to a factor of 10 within $\sim$100--600 days following the outbursts
\citep{akn+15}. The torque variations then abated and the torque
decreased to a relatively steady quiescent value. After both its
2008 and 2009 outbursts, \fift\ experienced a rapid increase in
spin-down torque \citep{dksg12}, and large fluctuations continue
(F.\ Camilo et al., in preparation). This behavior appears
broadly comparable to that displayed by \src\ until mid-2011
(Figure~\ref{fig:fdot_disc}).

The history of the transient magnetar \firstradio\ since its one
known X-ray outburst, detected in 2003, provides a more complete
parallel to the overall behavior exhibited by \src\ since 2009.
Following its outburst, \firstradio\ at first displayed erratic
variations in torque \citep{gh07}.  By 2006--2007, radio observations
showed a large but monotonic decrease in $|\dot\nu|$ \citep{ccr+07}.
The torque and radio flux density then (relatively) stabilized at
low values for approximately one year, after which the detectable radio emission
ceased \citep{crh+16}. On the whole, this seems to track
what we have observed in \src\ since 2009 (Figures~\ref{fig:fdot_disc}
and \ref{fig:fluxes}b).

\subsection{Twisted Magnetosphere Model}

In the twisted magnetosphere model for magnetar outbursts, the X-ray and radio 
emission are both caused by twists in the magnetosphere that can result from
shearing of the crust due to a starquake \citep{bel09}. The bundle of these
twisted, closed field lines is called a ``j-bundle''. The X-rays result from
a hotspot on the crust heated by currents driven by the j-bundle and the radio 
emission originates from the currents in the j-bundle itself.

\citet{bel09} apply their model to the outburst of \firstradio, an event
to which the behavior of \src\ displays some parallels.
The non-monotonic behavior in the spin-down torque observed for \firstradio\
is attributed by \citet{bel09}
to the increase of the twist angle at early times after the outburst. 
Depending on the initial conditions of the twist, the twist angle can
grow as the j-bundle is shrinking causing the poloidal field lines to 
inflate, opening them at the light cylinder. This increase of the magnetic field
at the light cylinder causes an increase in the spin-down torque. 
Once the twist angle reaches a maximum, the torque then decreases monotonically.
This picture broadly fits the observed spin-down of \src\ (Figure \ref{fig:fdot_disc}),
where an epoch of fluctuating torque after discovery was
followed by a smooth monotonic decrease in $|\dot \nu|$ during 2011--2014.

In this model, we may expect the width of the radio pulse profile
to decrease as the j-bundle shrinks. For \src, we do observe a
decrease in the width of the trailing pulse profile component prior
to the disappearance of detectable radio emission (Figure
\ref{fig:gauss_fits}).  However, the leading profile component
remained constant in width, and a shrinking j-bundle therefore does
not appear to entirely account for the observed evolution of the
pulse profile.

\section{Conclusions}
\label{sec:conc}

We have presented 5\,yr of new Parkes radio observations of the
magnetar \src. We find that the torque on the neutron star decreased
monotonically from late 2011 through 2014 March, decreasing at the
smallest rate ever observed for this object starting in early 2013.
The flux density, while variable, reached a relatively steady low
level starting in early 2013 as well. 
Along with these decreases in flux density and torque, the pulse profile
evolved in a secular fashion where the pulse became narrower as the secondary
component approached the leading component.
Sometime in the last 9 months
of 2014, radio emission ceased and remained undetectable as of late
2016. This broadly parallels the behavior of the first radio magnetar, the
transient \firstradio.

The huge and rapid torque variations displayed by \src\ during
2009--2011 \citep[][and Figure~\ref{fig:fdot_disc}]{lbb+12}, akin
to those shown by other magnetars within a couple of years of X-ray
outbursts \citep[e.g.,][]{akn+15}, together with its exponentially
decaying X-ray flux during 2007--2011 \citep{ags+12}, argue for an
undetected outburst occurring not long before mid-2007.  On the
other hand, the radio behavior of \src\ during 1999--2004 \citep[][and
Figure~\ref{fig:fluxes}]{lbb+10} suggests an unsettled magnetosphere
as far back as at least 1999.  We may therefore suppose that \src\
was not in quiescence for many years preceding its ``2007''
outburst.

Quite likely, \src\ is currently in as quiescent a state as it has
been since at least 1999.  This is supported by the smoothly
decreasing torque in 2013--2014 and turn off of any detectable radio
emission by 2015.  We continue to monitor for the re-activation of
radio pulsations with Parkes, although by analogy with other radio
detected magnetars this may not happen until a new X-ray outburst.
The current X-ray state of \src\ is unknown; a measurement of its
X-ray flux would be very useful both to compare it to other magnetars
and to provide a baseline for the next outburst that will surely
arise. \\

\acknowledgements

The Parkes Observatory is part of the Australia Telescope, which
is funded by the Commonwealth of Australia for operation as a
National Facility managed by CSIRO.
P.S. was supported by a Alexander Graham Bell Canada Graduate Scholarship from
NSERC and a Schulich Graduate Fellowship from McGill University. P.S. holds
a Covington Fellowship at DRAO.

\bibliographystyle{apj}
\bibliography{journals_apj,myrefs,modrefs_mcgill,modrefs,psrrefs,crossrefs}

\end{document}